\documentclass[reprint,twocolumn,superscriptaddress,aps]{revtex4-2}

\usepackage{color}
\usepackage{graphicx}
\usepackage{mathtools}
\usepackage{verbatim}
\usepackage{amsmath}
\usepackage[font=small]{caption}
\usepackage{subcaption}
\usepackage{appendix}
\usepackage{esint}
\usepackage{geometry}
\usepackage{lipsum}
\usepackage{amsfonts}
\usepackage{titlesec}
\usepackage{etoolbox}
\usepackage{float}
\usepackage[utf8]{inputenc}
\usepackage{booktabs}
\usepackage{tabularx}
\usepackage{colortbl}
\usepackage{dcolumn}
\usepackage{braket}
\usepackage{bm}
\usepackage{textcomp}
\usepackage{hyperref}
\usepackage{amssymb, physics, siunitx}

\begin{document}

\title{Dual Magnetic and Electric Dipole Symmetry: Pseudo-Angular Momentum in Parity Space and the Electric Land\'e $g$-Factor}

\author{Michael E. Tobar}
\affiliation{Quantum Technologies and Dark Matter Research Lab, Department of Physics, University of Western Australia, Crawley, WA 6009, Australia}


\begin{abstract}
Electric dipole moments (EDMs) are sensitive probes of fundamental symmetries and central to searches for physics beyond the Standard Model. We present a symmetry-based, Zeeman-analogue operator framework that places magnetic and electric dipole physics on parallel footing under electromagnetic duality, and introduce a polar-sector pseudo-angular-momentum degree of freedom in parity space together with an associated electric Land\'e factor that organizes induced orbital dipoles. Following Ohanian's effective-current formulation of the Zeeman effect, we construct its electric dual: the wavefunction's microscopic polarization admits an equivalent effective magnetic probability-current representation, providing a field-equivalence description of parity-mixed charge displacement. In this notation the total EDM expectation takes the unified form $\langle \hat{\vec d}_{\rm tot}\rangle= d_B(g_E\,\frac{\langle \hat{\vec J}_p\rangle}{\hbar}+ g_E^{e}\,\frac{\langle \hat{\vec S}\rangle}{\hbar})$, with $g_E^{e}=\frac{2d_{\rm int}}{d_B}$, where $\hat{\vec J}_p$ captures Stark-induced pseudo-angular momentum and $\hat{\vec S}$ encodes any intrinsic (spin-aligned) EDM $d_{\rm int}$ from symmetry-violating interactions. We define a natural electric dipole unit (the ``Bohr EDM'') as $d_B \equiv e a_0=\frac{2\mu_B}{c\alpha}$ ($a_0$ the Bohr radius and $\mu_B$ the Bohr magneton). As a canonical analytic benchmark, we show in the hydrogenic problem that a static electric field couples within a fixed $n$ manifold through the scaled Runge--Lenz structure, yielding a compact Land\'e-like description and reproducing the Stark doublet (e.g.\ $|\langle d_{\rm orb}\rangle|=3d_B$ for the $2s$--$2p_{m=0}$ mixing).

\end{abstract}


\maketitle

\section{Introduction}
Electric dipole moments (EDMs) are exceptionally sensitive probes of fundamental symmetries and play a central role in the search for physics beyond the Standard Model. In atomic and molecular systems, EDMs can arise through two distinct mechanisms: as \emph{intrinsic EDMs}, aligned with the particle’s real spin and signaling new CP-violating physics, or as \emph{induced EDMs}, generated by parity mixing of orbital states and as we show in this paper, may naturally be described by a pseudo-angular-momentum degree of freedom, as in the linear Stark effect. Although these mechanisms are conceptually different, one tied to spin in position space, the other to pseudo-angular-momentum in parity space, both manifest experimentally as level shifts under external electric fields, a signature exploited in high-precision spectroscopy and EDM searches.

The linear Stark effect is a well-known example of an EDM arising from orbital dynamics, and is the electric analogue of the Zeeman effect. This phenomenon describes the splitting or shifting of atomic energy levels in a static electric field \cite{Stark1914}. For a nonzero first-order shift to occur, the system must contain states of opposite parity that are sufficiently degenerate. In excited hydrogen, for instance, the $n=2$ manifold includes the degenerate $2s$ and $2p$ orbitals. An applied electric field mixes these states, distorting the electron cloud and inducing a dipole moment aligned with the field, which breaks inversion symmetry and couples to the orbital motion, producing a first-order energy shift \cite{Feynman57,EDMBook}.

In contrast, EDMs associated with intrinsic spin are expected to be permanent and arise from violations of discrete symmetries, particularly parity (P) and time reversal (T) invariance. Within quantum field theory, such EDMs emerge from CP-violating operators and are predicted to couple directly to the spin vector via terms such as,
\begin{equation}
\mathcal{L}_{\text{EDM}} = -\frac{1}{2} d_{int}\bar{\psi} \sigma^{\mu\nu} \gamma^5 \psi F_{\mu\nu}.
\label{Lint}
\end{equation}
Here, $d_{int}$ is the intrinsic EDM of the particle (in units of $C\cdot m$), $\psi$ is the Dirac spinor field, $\sigma^{\mu\nu} = \frac{i}{2} \gamma^\mu, \gamma^\nu $ is the antisymmetric spin tensor, $F_{\mu\nu}$ is the electromagnetic field strength tensor and $ \gamma^5$, ensures the interaction violates both parity (P) and time-reversal (T) symmetry. The detection of a spin-aligned EDM in a fundamental particle like the electron or neutron would constitute direct evidence of physics beyond the Standard Model and provide insights into the baryon asymmetry of the universe.

Despite decades of effort, no intrinsic EDM has been observed to date. Current experiments place stringent limits on possible CP-violating effects\cite{Sandars1965,Engel2013,DeMille2017,Chupp2019,Regan2002}, driving continued interest in EDM searches across atomic, molecular, and solid-state platforms. Theoretical frameworks beyond the Standard Model such as supersymmetry, left-right symmetric models, and theories of baryogenesis often predict EDMs just below current experimental sensitivity.

Complementary to these quantum approaches, insights from condensed matter and electromagnetic energy generation provide an alternative view of EDM-like phenomena. For example, modern polarization theory models permanent electric polarization in materials using the Berry phase formalism, where each unit cell contributes a quantized dipole moment \cite{Resta1994,Vanderbilt2018,Song2021}. This macroscopic charge separation is analogous to intrinsic EDMs in fundamental systems. Active voltage sources also create permanent polarizations \cite{TobarVP,Tobar2021,TobarModAx19}, and reflect symmetry-breaking mechanisms similar in form to those studied in particle physics\cite{Hwang2012,Liu2020}.

In this work we build on the magnetic-electric analogy to develop a semi-classical model of EDMs, inspired by the Bohr magneton and by Ohanian’s analysis \cite{Ohanian86}. Ohanian established a connection between magnetic moment and electron spin by showing that spin can be viewed as arising from a circulating flow of energy in the wave field, analogous to the angular momentum carried by a classical circularly polarised wave. In doing so, he introduced an effective probability current density, analogous to a bound current in a macroscopic magnetised medium, defined as the curl of an effective microscopic magnetisation field constructed directly from the electron wavefunction.

By electromagnetic duality, we construct an electric analogue of Ohanian’s picture: we introduce a “Bohr EDM” and a dual-current description in which an effective probability magnetic current arises from the curl of a microscopic polarisation field constructed from the wavefunction. In this framework, induced EDMs, such as those of the Stark-mixed hydrogen doublet, are organised by a polar-sector pseudo-angular-momentum in parity space (with an explicit Runge–Lenz connection in the hydrogenic case). This pseudo-angular momentum is not a physical rotation in space, but a measure of parity mixing that determines the induced polarisation. The same Zeeman-analogue operator language accommodates both induced orbital EDMs (from parity mixing/charge displacement) and intrinsic spin-aligned EDMs within a unified framework; hydrogen is used as the canonical analytic benchmark where the algebra can be demonstrated in closed form, so the familiar Stark eigenstates are reproduced by construction while the main contribution is the underlying duality-driven operator structure and its physical interpretation.

\section{Magnetic Dipole Moments}

\subsection{The Semi-Classical Bohr Magneton}

The Bohr magneton arises naturally from a semi-classical model of an electron orbiting a nucleus, as in the Bohr model of the atom. In this picture, an electron of charge $- e$ and mass $m_e$ moves in a circular orbit of radius $r$ with speed $v$, generating a current, $I$.
\begin{equation}
I=\frac{e}{T}=\frac{e v}{2 \pi r}
\end{equation}
To draw a classical parallel, consider a loop of current $I$ encircling an area $\vec{A}$. The magnetic dipole moment associated with this loop is given by:
\begin{equation}
\vec{\mu} = I \vec{A}.
\end{equation}
For an electron in a circular orbit of radius $r$ and speed $v$, the current is $I = e/T = ev/(2\pi r)$, and the area is $A = \pi r^2$, yielding:
\begin{equation}
\mu = \frac{e v r}{2} = \frac{e L}{2m_e},
\end{equation}
where $L = m_e v r$ is the orbital angular momentum. 

In the Bohr model, the quantisation condition imposes $L=\hbar$, for the lowest energy level, which yields a magnetic moment
\begin{equation}
\mu _B=\frac{e\hbar}{2m_e},
\label{BMag}
\end{equation}
defining the Bohr magneton. This derivation illustrates how the calassical concept of a circulating charge loop, combined with quantum angular momentum quantisation, leads to a fundamental unit of magnetic moment in atomic physics.

\subsection{Zeeman Effect due to Intrinsic and Orbital Magnetic Moments}

In the presence of an external static and uniform magnetic field $\vec B$, the perturbation Hamiltonian arises from the interaction of the electron magnetic moments with $\vec B$.
\begin{align}
\hat{H}_Z=-\hat{\vec{\mu}}\cdot\vec{B}
\end{align}
where the magnetic moment, $\vec{\mu}=\vec{\mu}_L +\vec{\mu}_S$, the sum of the orbital and intrinsic spin magnetic moments.
\begin{equation}
\begin{aligned}
\hat{\vec{\mu}}_L &= -\frac{\mu_B}{\hbar} \hat{\vec{L}}, \\
\hat{\vec{\mu}}_S &=-g_S\frac{\mu_B}{\hbar} \hat{\vec{S}}.
\label{ZM}
\end{aligned}
\end{equation}
Here $g_S$ is the electron spin $g$-factor, which in Dirac theory equals 2, and in QED $\simeq 2.0023192$. While orbital moments follow from classical current loops, the spin moments are considered to have no classical counterpart.

This perturbation leads to lines splitting into several components in the presence of the magnetic field, with Zeeman corrections characterised by,
\begin{equation}
\hat{H}_M=g_M\frac{\hat{J}_z}{\hbar}\mu_BB_z; ~ \Delta E_M=g_Mm_j\mu_BB_z,
\label{mgf}
\end{equation}
where $\hat{H}_M$ is the the Zeeman Hamiltonian, $\hat{J}_z$ is the z-component of the total angular momentum, $\hat{\vec{J}}$, while, $\Delta E_M$ is the energy shift of each Zeeman sub-level, $g_M$ is the magnetic Land\'e g-factor, and $m_j= -J,-J+1...,J$, is the magnetic quantum number associated with the total angular momentum projection.

\subsection{Ohanian's Model of Spin: Magnetic Moment from Internal Magnetisation}

In his seminal paper \textit{``What is Spin?''}~\cite{Ohanian86}, Ohanian provides a physically motivated description of the electron's magnetic moment by modeling it in terms of an effective probability quantum current distribution, despite the fact that spin has no classical analog in terms of rotating charge. His treatment shows that the electromagnetic properties of spin, specifically the magnetic moment, can be derived from a localised magnetisation density constructed from the Dirac wavefunction.

Ohanian introduces a microscopic magnetization vector field $\vec{M}$ defined as
\begin{equation}
\vec{M} = \mu_B \psi^\dagger \vec{\Sigma} \psi,
\end{equation}
where $\psi$ is the Dirac spinor, and $\vec{\Sigma}$ represents the spin matrices. This magnetization leads to an effective bound probability current density,
\begin{equation}
\vec{J}_b = \nabla \times \vec{M},
\end{equation}
analogous to classical magnetization currents. The magnetic moment $\vec{\mu}$ associated with this probability current is:
\begin{equation}
\vec{\mu} = \frac{1}{2} \int \vec{r} \times \vec{J}_b \, d^3r.
\end{equation}
Using the relation to magnetization:
\begin{equation}
\vec{\mu} = \int \vec{M} \, d^3r,
\end{equation}
Ohanian recovers the quantum result:
\begin{equation}
\hat{\vec{\mu}}_S=-g_S \frac{\mu_B}{\hbar} \hat{\vec{S}}, \quad g_S = 2.
\end{equation}

Ohanian's construction reveals that although spin cannot be interpreted as literal circulating charge, its associated magnetic moment behaves as if it originates from an effective probability current density. This provides a bridge between classical electromagnetism (magnetization and bound currents) and quantum field theory (spinor structure and symmetry). The result reinforces the idea that the magnetic moment of a spin-$ \frac{1}{2} $ particle is not merely a formal operator feature, but corresponds to a real, observable electromagnetic quantity tied to internal degrees of freedom in the quantum field.

\section{Electric Dipole Moments}

So far, no particle has been observed to have a nonzero intrinsic EDM, and are predicted to exist but at levels far below current experimental sensitivity (for example, the electron EDM is expected to be about $10^{-38}\,\mathrm{e\cdot cm}$, whereas the neutron is about $10^{-31}\,\mathrm{e\cdot cm}$). State-of-the-art experiments constrain the electron EDM to be smaller than $4.1 \times 10^{-30}\,\mathrm{e\cdot cm}$ \cite{Roussy23} and the neutron EDM below $1.8 \times 10^{-26}\,\mathrm{e\cdot cm}$ \cite{Abel2020}, with similar null results for the muon and proton. These increasingly tight bounds are crucial because a confirmed intrinsic EDM would signal new CP-violating physics beyond the Standard Model, possibly helping to explain the matter-antimatter asymmetry of the Universe. By contrast, the electric dipole moments that appear in the linear Stark effect of hydrogen or polar molecules are \emph{not intrinsic but induced}. For example, in hydrogen they arise from parity mixing of nearly degenerate orbitals, such as the $2s$ and $2p_0$ states. In this case, the dipole operator acts on a \emph{pseudo-angular momentum degree of freedom}, defined in the two-level parity subspace, and the resulting EDM depends on the external field and the mixing angle. 


\begin{figure}[t]
\includegraphics[width=1.0\columnwidth]{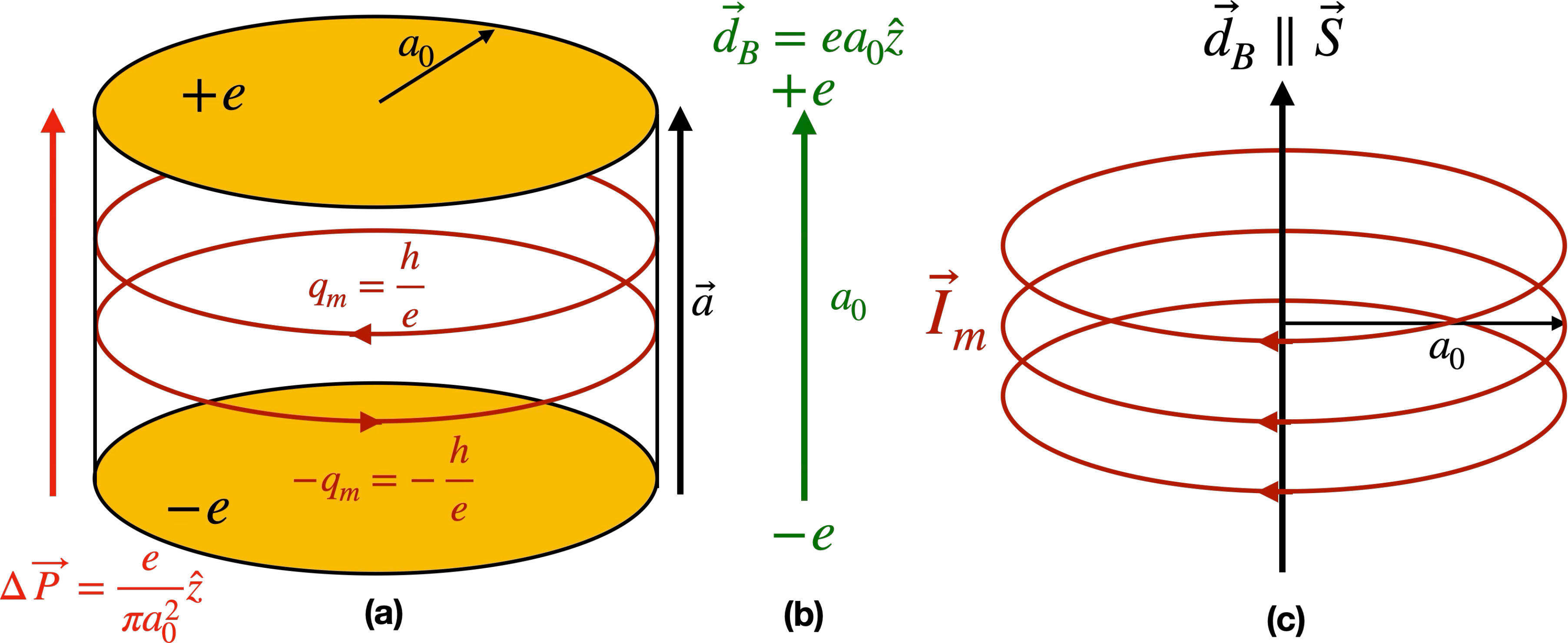}
\caption{Semi-classical representations of the Bohr EDM scale, $\vec{d}_B$.
(a) Idealized quasi-1D “unit cell” of quantized permanent polarization,
$\Delta\vec{P}=\frac{e}{\pi a_0^2}\hat{z}$, for which $\nabla\times\vec{P}\neq 0$ implies an \emph{effective} magnetic current on the radial surface (a dual representation of the polarization texture, not a literal magnetic-charge flow).
For mnemonic purposes this surface current can be pictured as two fictitious opposite magnetic charges $\mp q_m=\mp\frac{h}{e}$ circulating in opposite senses, associated with pseudo-spin (pseudo-angular momentum) in parity space rather than mechanical orbital motion.
(b) Classical Bohr EDM: two point charges $\pm e$ separated by the Bohr radius $a_0$.
(c) Schematic of an intrinsic Bohr-scale EDM connected to the position-space spin vector $\vec S$, represented equivalently by an effective azimuthal magnetic current $\vec{I}_m$ at radius $\sim a_0$.}
\label{UC}
\end{figure}

For atomic scale, we define the atomic dipole unit, the Bohr EDM as
\begin{equation}
d_B \;\equiv\; e a_0,
\end{equation}
In the Coulomb problem this scale is consistent with the hydrogenic $SO(4)$ (Runge-Lenz) structure within a fixed $n$-manifold, and it also has an immediate classical interpretation: it is the dipole moment of two point charges $\pm e$ separated by a distance $a_0$ [Fig.~\ref{UC}(b)]. Furthermore, the same dipole unit appears in a quasi-1D “cell” formulation governed by modern polarization theory, when the size of the cell spacing is of order $a_0$.

The following toy model is included only as a \emph{heuristic mnemonic} (not as a physical derivation) to connect this dipole scale to a dual-current picture. Similar to the semi-classical Bohr magneton estimate, we imagine an effective surface magnetic current associated with the curl of the polarization. For mnemonic purposes, this may be visualized as two fictitious opposite “magnetic charges” $\pm \frac{h}{e}$ circulating at the Bohr radius $a_0=\frac{\hbar}{m_{\mathrm{e}} c \alpha}$, where $\alpha=\frac{e^2}{4 \pi \varepsilon_0 \hbar c}$.

In this picture, the net electric charge, net magnetic charge, and total angular momentum of the cell are zero, but an effective magnetic surface current exists, which we associate with pseudo-spin (parity mixing) rather than mechanical orbital motion. With $q_m=\frac{h}{e}$, the effective magnetic current at the boundary is written as
\begin{equation}
\vec{I}_{m}=-\frac{2q_m v}{2\pi a_0 } \hat{\phi}
= -\frac{e}{2\pi a_0^2m^*\epsilon_0\alpha c} (a_0m^*v)\hat{\phi},
\end{equation}
where $m^*$ is an effective inertia parameter for the internal (parity-space) degree of freedom. For a Bohr-scale dimensional estimate we take the usual quantization scale $a_0 m^* v \sim \hbar$ with $m^*\sim \mathcal{O}(m_e)$ (the precise value of $m^*$ is not used elsewhere in the manuscript). This gives
\begin{equation}
\vec{I}_m \approx -\frac{e\hbar}{\pi a_0^2m_e\epsilon_0\alpha c} \hat{\phi}
= -\frac{2\mu_{\mathrm{B}}}{\pi a_0^2\epsilon_0\alpha c} \hat{\phi},
\label{magcur}
\end{equation}
and we may express the Bohr EDM in terms of the effective azimuthal magnetic current (left-hand rule) as
\begin{align}
d_B=-\epsilon_0I_m\pi a_0^2=\frac{2\mu_{\mathrm{B}}}{c\alpha}=e a_0,
\label{BElec}
\end{align}
the electromagnetic dual of the Bohr magneton in (\ref{BMag}). Thus Fig.~\ref{UC}(a) and Fig.~\ref{UC}(b) represent the same Bohr-scale dipole unit: $\vec a=a_0\hat z$ implies $d_B=e a_0$.


In modern polarization theory, the bulk polarization is defined “modulo a quantum”, along with a Berry phase defined modulo $2\pi$, and is given by
\begin{equation}
\Delta \vec P \;=\; \frac{e\,\vec a}{V},
\label{eq:pol-quantum-3D}
\end{equation}
where $\vec a$ is any lattice vector and $V$ the unit-cell volume. This follows from an electron in a periodic potential, which gives the $U(1)$ gauge freedom (Bloch's theorem); $\psi_{n\vec k}\!\to\!e^{-i\vec k\!\cdot\!\vec a}\psi_{n\vec k}$, which shifts the Berry phase by $2\pi$ and $\,\vec P$ by $e\vec a/V$. Here we consider a quasi-1D “cell” as shown in Fig.~\ref{UC} of length $a$ (along $\hat{\vec z}$) and cross-section $A_{\!\mathrm{eff}}$, so the effective volume is $V_{\!\mathrm{eff}}=A_{\!\mathrm{eff}}a$. In our construction $2\pi$ winding of the internal phase is equivalent to choosing $\vec a=a\hat{\vec z}$. Then from \eqref{eq:pol-quantum-3D}
\begin{equation}
\Delta P_z \;=\; \frac{e\,a}{V_{\!\mathrm{eff}}}
\;=\; \frac{e}{A_{\!\mathrm{eff}}}
\;=\; \frac{e}{\pi a_0^2}.
\label{eq:DeltaP_e_over_A}
\end{equation}
This is equivalent to a single cell of a supercell in modern polarization theory \cite{Vanderbilt2018}, with a quantum of permanent polarization $\Delta P_z$ set by $A_{\!\mathrm{eff}}=\pi a_0^2$, as shown in Fig.~\ref{UC}. With $\vec{a}=a_0\hat{z}$, the dipole moment follows as $\vec{d}_B=\int \Delta P_z\, d^3r=ea_0\hat{z}=d_B\hat{z}$, consistent with (\ref{BElec}).

For completeness when discussing the hydrogenic problem, we note that, for a finite-mass nucleus, the mass parameter in the internal (relative-coordinate) Coulomb problem should be understood as the reduced mass
\begin{equation}
\mu \;=\; \frac{m_e m_N}{m_e+m_N},
\end{equation}
rather than the electron mass $m_e$, so that the Bohr radius is replaced by
\begin{equation}
a_\mu \;=\; \frac{4\pi\epsilon_0\hbar^2}{\mu Z e^2},
\end{equation}
and hence the Bohr dipole scale used throughout becomes $d_B = e a_\mu$ (rather than $e a_0$ defined using $m_e$). This substitution rescales lengths and energies but does not modify the underlying symmetry structure: the $SO(4)\simeq SU(2)\times SU(2)$ algebra and associated commutation relations remain unchanged (see Appendix C), with $m\rightarrow\mu$ in the Hamiltonian. In practice, for hydrogen the correction is very small,
\begin{equation}
\frac{\mu}{m_e} \;=\; \frac{1}{1+m_e/m_p} \;\approx\; 0.99946,
\end{equation}
so $a_\mu$ (and thus $d_B$) differs from the $m_e$ value by only $\sim 5.4\times 10^{-4}$ (about $0.054\%$), and either convention may be adopted depending on the desired level of precision. For clarity, we emphasise that throughout the paper we use the internal (relative-coordinate) formulation of the two-body problem, and therefore all ``densities/fields'' refer to quantities defined in the nucleus/centre-of-mass frame.

\section{Dual Ohanian Model: Dipole moment from Internal Polarisation}
\label{Oh}

\subsection{Bohr Electric Dipole Moment from Spin}

As discussed previously, In Ohanian's seminal work, the intrinsic magnetic moment of the electron was interpreted semiclassically as arising from an effective circulating energy flow of the electron wavefunction, generating an angular momentum and associated magnetic dipole \cite{Ohanian86}. The effective magnetization density was defined from the bilinear spinor quantity $\psi^\dagger \vec{\Sigma} \psi$, leading to the construction of an effective probability current density $\vec{J}_b = \nabla \times \vec{M}$, from which the Bohr magneton,  $\mu_B$, naturally arises. 

In analogy, we pose a putative dual picture in which a permanent EDM arises from an internal effective probability magnetic current distribution, that creates a permanent polarization by a left hand rule. This provides a complementary viewpoint in which the EDM can be understood as the macroscopic manifestation of internal field circulations analogous to those producing magnetic dipole moments. Thus, by invoking an analogue dual construction, one may define a spin operator, $\hat{\vec S}$, associated with the following internal microscopic polarization density,
\begin{align}
  \vec{P}(\vec{r}) &=d_B\,\psi^\dagger(\vec{r})\,\vec{\Sigma}\,\psi(\vec{r}),\qquad d_B \equiv e a_0, \\
  &= \frac{2d_B}{\hbar}\psi^\dagger(\vec{r})\hat{\vec S}\psi(\vec{r}),\qquad \hat{\vec S}\equiv \frac{\hbar}{2}\vec{\sigma}. 
\end{align}
where, $\vec{\sigma}$ denotes the vector of Pauli operators acting on the two-level parity space, $\vec{\Sigma}$ is the Pauli spin matrix vector and $\psi$ the Dirac or Pauli spinor.  Adopting the dual (Weber) convention for magnetic current, the curl of $\vec{P}$ defines an effective probability \emph{magnetic} current density,
\begin{equation}
  \vec{J}_m \;=\; -\,\frac{1}{\epsilon_0}\nabla\times\vec{P},
  \label{MCdens}
\end{equation}
which plays the role dual to Ohanian's effective current from $\vec{M}$. Since, $\|\langle\vec{\Sigma}\rangle\|=1$, the expectation value of the electric dipole moment is then
\begin{equation}
 \langle\vec{d}\rangle=\int \vec{P}(\vec{r})\,d^3r \;=\; d_B,
  \label{dPol1}
\end{equation}
consistent with
\begin{equation}
\hat{\vec{d}}=\frac{g_E^Sd_B}{\hbar}\hat{\vec S} ~~g_E^S=2,
\end{equation}
where $g_E^S$ is the electric Land\'e g-factor. Thus, a fully polarised single-particle state carries the ``Bohr'' electric dipole moment $d_B=e a_0$. This gives rise to an effective electric dipole moment associated with the spin degree of freedom. The spin carries an intrinsic dual moment generated by the internal microscopic polarization of the wavefunction. This construction provides a dual Ohanian model that allows a semiclassical interpretation of an effective intrinsic electric dipole moment.

In this framework, the magnetic moment $\vec{\mu}$ due to a circulating electric current is mirrored by an electric dipole moment $\vec{d}$ due to a circulating magnetic current:
\begin{equation}
 \langle\vec{d}\rangle = \frac{\epsilon_0}{2} \int \vec{r} \times \vec{J}_m \, d^3r,
  \label{dPol2}
\end{equation}
with a semi-classical picture shown in Fig.\ref{UC}(c). This construction preserves the correct transformation properties under parity (P) and time-reversal (T), making it consistent with known symmetry-breaking signatures of EDMs. However, in reality no EDMs coupled to intrinsic spin have yet to been observed, and if they do exist are significantly smaller than Bohr EDM.

\subsection{Intrinsic and Orbital Polarization}

If the electron carries an intrinsic electric dipole moment, $d_{int}$, the local microscopic polarization density of the electron wavefunction would be of the form,
\begin{equation}
\vec{P}_{\rm e}(\vec{r}) = d_{int}\,\psi^\dagger(\vec{r})\,\vec{\Sigma}\,\psi(\vec{r})
\;\;\approx\;\; \frac{2 d_{int}}{\hbar}\,\vec{s}(\vec{r}),
\end{equation}
consistent with
\begin{equation}
\hat{\vec{d}}_e=\frac{g_E^ed_B}{\hbar}\hat{\vec S} ~~~~g_E^e=\frac{2d_{int}}{d_B},
\label{espin}
\end{equation}
where $\vec{s}(\vec{r})=\psi^\dagger(\vec{r})\frac{\hbar}{2}\vec{\sigma}\psi(\vec{r})$ is the nonrelativistic spin density and $g_E^e$ is the effective electric Land\'e g-factor of the electron. To understand how the electron behaves in an electric field we consider the interaction term in (\ref{Lint}). To interpret this interaction physically, we examine its nonrelativistic limit, and  begin by noting the electromagnetic field strength tensor:
\begin{equation}
F^{\mu\nu} =
\begin{pmatrix}
0 & -E_x & -E_y & -E_z \\
E_x & 0 & -B_z & B_y \\
E_y & B_z & 0 & -B_x \\
E_z & -B_y & B_x & 0
\end{pmatrix}.
\end{equation}

The term $ \bar{\psi} \sigma^{\mu\nu} \gamma^5 \psi F_{\mu\nu} $ contracts to:
\begin{equation}
2\bar{\psi} \sigma^{0i} \gamma^5 \psi F_{0i} + \bar{\psi} \sigma^{ij} \gamma^5 \psi F_{ij},
\end{equation}
where the dominant term at low velocities is the electric coupling:
\begin{equation}
\bar{\psi} \sigma^{0i} \gamma^5 \psi \approx \psi^\dagger \Sigma^i \psi, \qquad F_{0i} = -E^i.
\end{equation}
Thus,
\begin{equation}
\bar{\psi} \sigma^{\mu\nu} \gamma^5 \psi F_{\mu\nu} \to -2 \, \psi^\dagger \vec{\Sigma} \psi \cdot \vec{E}.
\end{equation}
Substituting back into the Lagrangian (\ref{Lint}), we obtain:
\begin{equation}
\mathcal{L}_{\text{EDM}} \to d_{int} \, \psi^\dagger \vec{\Sigma} \psi \cdot \vec{E} = - \vec{P}(\vec{r}) \cdot \vec{E},
\end{equation}
which corresponds to the energy density term.

Generalizing to an atomic state $\psi_\pm$, the total polarization may be expanded to the sum of the orbital (charge-displacement) part and the EDM (spin) part,
\begin{equation}
\vec{P}_{\rm tot}(\vec{r})
= \underbrace{\rho_{\rm charge}(\vec{r})\,\vec{r}}_{\text{orbital (Stark)}} \;+\;
\underbrace{d_{int}\,\psi_\pm^\dagger(\vec{r})\,\vec{\Sigma}\,\psi_\pm(\vec{r})}_{\text{spin EDM}}.
\label{Poltot}
\end{equation}
Integrating over space gives the total permanent dipole moment
\begin{align}
&\langle\vec{d}_{\rm tot}\rangle
= \int \vec{P}_{\rm tot}\,d^3r
= \langle\vec{d}_{\rm orb}\rangle \;+\; d_{int}\,\langle \vec{\Sigma}\rangle_{\psi_\pm}, \\
&\langle\vec{d}_{\rm orb}\rangle\simeq\int \rho_{\rm charge}(\vec{r})\,\vec{r}\,d^3r.
\end{align}
Since no electron EDM has been observed, it either vanishes or the intrinsic value $d_{\rm int}$ lies below current experimental sensitivity. Throughout this work we set $d_{\rm int}\simeq 0$, so that the polarization obtained from (\ref{Poltot}) reduces to $\vec P(\vec r)\simeq \rho_{\rm charge}(\vec r)\,\vec r$. To calculate orbital EDMs (e.g., the linear Stark effect), we use the effective magnetic probability current density, Eq.~(\ref{MCdens}), together with the polarization relations (\ref{dPol1}) and (\ref{dPol2}). Importantly, the induced orbital EDM is produced physically by parity mixing (charge displacement); the associated magnetic-current description is a topological field equivalence (a dual representation of the polarization texture), not an additional causal mechanism. We further show that this charge displacement can be organized in terms of an \emph{effective} angular-momentum like degree of freedom, a pseudo-angular momentum in parity space, that emerges when the system possesses an internal two- (or few-) level structure governed by a hidden symmetry.

\section{The Stark Effect for the Hydrogen Atom}

To make the analogy between magnetic and electric dipole moments concrete, we consider the classic example of the linear Stark effect in the hydrogen atom. The perturbing Hamiltonian is
\begin{equation}
H_S = -\vec{d} \cdot \vec{E} = -q \vec{r} \cdot \vec{E},
\end{equation}
where $\vec{d} = q \vec{r}$ is the electric dipole operator and $\vec{E}$ is the applied electric field. For a field along the $z$-direction, $\vec{E}=E_z\hat{z}$, this becomes $H_S = -q z E_z$.

In first-order perturbation theory, the energy shift is
\begin{equation}
\Delta E_S^{(1)} = \bra{n,l,m} H_S \ket{n,l,m},
\label{En2}
\end{equation}
which vanishes unless the state has mixed parity. In this represenatation, the electron’s orbital quantum numbers are the principal quantum number $n=1,2,3,\ldots$, the orbital angular momentum $\ell=0,1,\ldots,n-1$ (with spectroscopic labels $s,p,d,f,\ldots$ for $\ell=0,1,2,3,\ldots$), and the magnetic quantum number $m=-\ell,-\ell+1,\ldots,\ell$, which set $L^2=\ell(\ell+1)\hbar^2$ and $L_z=m\hbar$.

In a degenerate subspace of opposite parity states, such as the $n=2$ manifold in hydrogen, the electric field couples $2s$ ($\ell=0$)and $2p$ ($\ell=1$) orbitals, leading to level splitting. Diagonalizing the perturbation in this subspace gives a permanent induced dipole moment and energy shifts linear in $E_z$.

This behavior mirrors the Zeeman effect, where magnetic dipole moments couple to an applied magnetic field. In both cases, an external field breaks an internal symmetry (parity or time-reversal), revealing hidden structure and inducing a dipole aligned to the field.

\subsection{Stark Effect for the $n=2$ Orbital}

\begin{figure}[t]
\includegraphics[width=1.0\columnwidth]{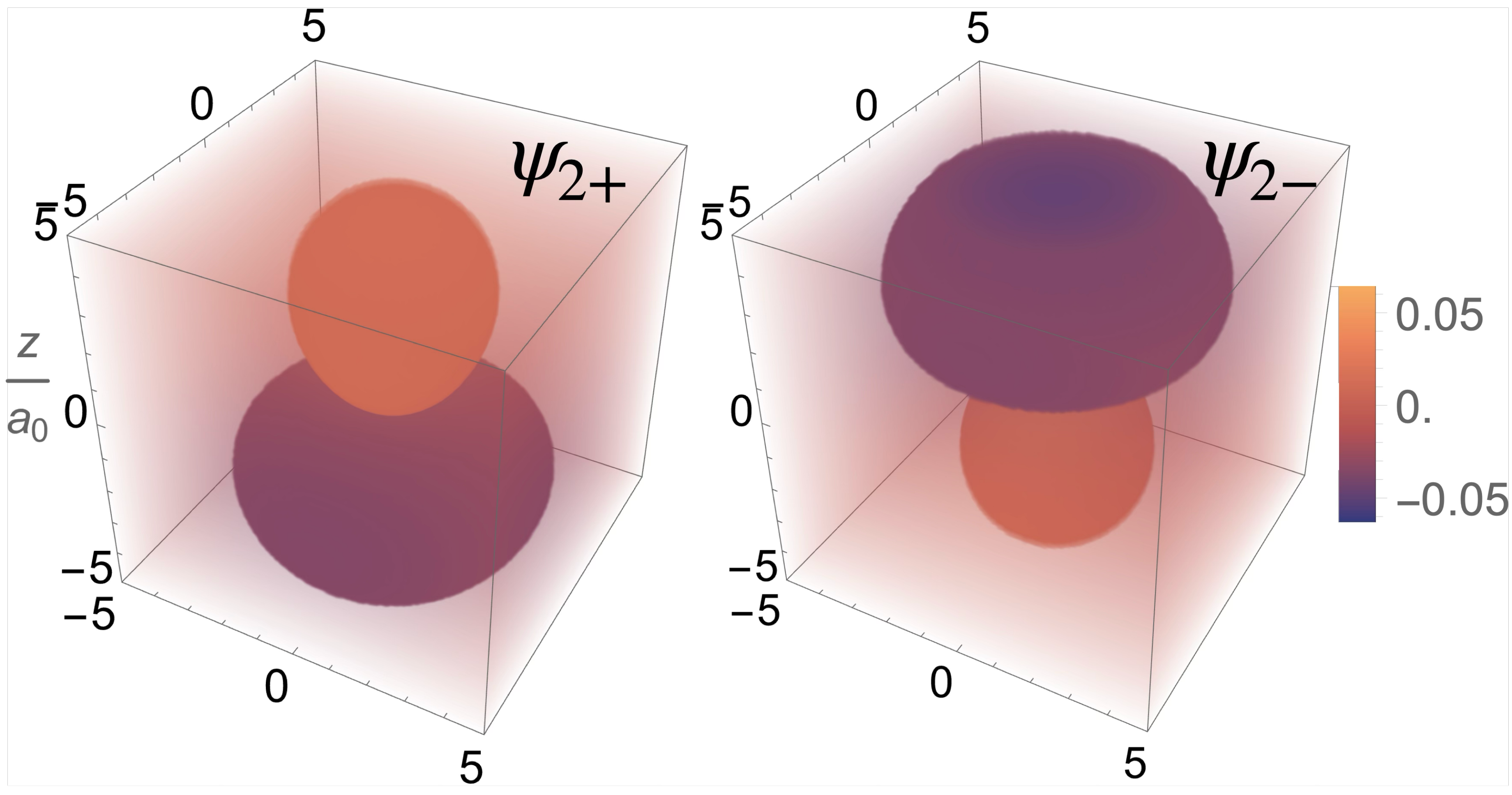}
\caption{Density plot of the Stark state wavefunctions, $\psi_{2\pm}$. The asymmetry of the electron cloud around the nucleus generates an electric dipole moment.}
\label{Psipm}
\end{figure}

The hydrogen eigenstates $\psi_{n\ell m}$ separate into a radial part $R_{n\ell}(r)$ and an angular part $Y_{\ell m}(\theta,\phi)$ in spherical coordinates $(r,\theta,\phi)$ by separation of variables in the Schr\"odinger equation with the Coulomb potential $V(r)=-\frac{e^2}{4\pi\epsilon_0 r}$. Explicitly,
\begin{equation}
\psi_{n\ell m}(r,\theta,\phi) = R_{n\ell}(r) \, Y_{\ell m}(\theta,\phi),
\label{WFn2}
\end{equation}
Here, we focus on the $n=2$ degenerate manifold which includes the following states:  The 2s state, $\psi_{200}$, and the 2p, $\psi_{21m}$ states, with $m = -1, 0, +1$. From (\ref{En2}) only the matrix element between $\psi_{200}$ and $\psi_{210}$ contributes at first order, given by,
\begin{equation}
    \matrixel{2s}{H_S}{2p_0} = -eE_z \matrixel{2s}{z}{2p_0}
\end{equation}
The normalised radial wave functions given in (\ref{WFn2}) for $n=2$ are known to be;
\begin{align}
    R_{20}(r) &= \frac{1}{2\sqrt{2}\,a_0^{3/2}}\left( 2 - \frac{r}{a_0} \right) e^{-r/(2a_0)}\\
    R_{21}(r) &= \frac{1}{2\sqrt{6}\,a_0^{3/2}}\left( \frac{r}{a_0} \right) e^{-r/(2a_0)},
\end{align}
and the angular wavefunctions to be;
\begin{align}
    Y_{00}(\theta,\phi) &= \frac{1}{\sqrt{4\pi}}, \\
    Y_{10}(\theta,\phi) &= \sqrt{\frac{3}{4\pi}} \cos \theta
    \label{angle}
\end{align}
so
\begin{align}
\psi_{2s}(r,\theta)   &= R_{20}(r)\,Y_{00} = \frac{R_{20}(r)}{\sqrt{4\pi}},\\
\psi_{2p_0}(r,\theta) &= R_{21}(r)\,Y_{10} = \sqrt{\frac{3}{4\pi}}\,R_{21}(r)\cos\theta.
\end{align}
The integral splits into radial and angular parts:
\begin{equation}
\begin{aligned}
   \matrixel{2s}{z}{2p_0}  &= \int \psi_{200}^*(\vec{r}) \, z \, \psi_{210}(\vec{r}) \, d^3r = \\
    &\left(\int_0^\infty R_{20}(r) r^3 R_{21}(r) \, dr\right) \cdot \\
    &\left(\int Y_{00}^* \cos\theta \, Y_{10} \, \sin\theta \, d\theta d\phi\right).
\end{aligned}
\label{Intl}
\end{equation}
Using the known integrals:
\begin{align}
    \int_0^\infty R_{20}(r) \, r^3 \, R_{21}(r) \, dr &= -3a_0 \\
    \int_0^{2\pi} d\phi \int_0^\pi Y_{00}^* \cos\theta \, Y_{10} \sin\theta \, d\theta &= 1,
\end{align}
we find,
\begin{equation}
    \matrixel{2s}{z}{2p_0} = -3a_0 
\end{equation}
Given the direction of the electric field, $\vec{E}$, defines the $z$-direction, The perturbation matrix in the $\{2s, 2p_0\}$ basis is:
\begin{equation}
    H = \begin{pmatrix} 0 & 3eE_za_0 \\ 3eE_za_0 & 0 \end{pmatrix}
\end{equation}
Diagonalizing this into the hybridised Stark basis, 
\begin{equation}
\psi_{2\pm}(r,\theta)=\frac{1}{\sqrt{2}}\Big[\psi_{2 s}(r,\theta)\pm\psi_{2 p_0}(r,\theta)\Big].
\label{hyb}
\end{equation}
yields the well known $n=2$ first-order energy Stark shifts,
\begin{equation}
    \Delta E^{(1)}_S = \pm 3 e E_z a_0 = \pm 3 E_z d_B = \pm E_z d_{2},
\end{equation}
where $d_{2}$ is the magnitude of the dipole moment, with the wavefunctions given by (\ref{hyb}), and plotted in Fig.\ref{Psipm}. A comparison of parameters for the $n=2$ and $n=3$ Stark states is shown in Tab. \ref{TabI}.

\subsection{Dual Ohanian Model for the $n=2$ Orbital}

\begin{table*}
\caption{Calculated values for $n=2,3$, where the Stark-states have non-zero $k$ values}
\begin{center}
\begin{tabular}{c|c|c|c|c|c}
$n$ & $g_E(n) = \frac{3}{2}n$ & $k$ values & $\displaystyle \int \frac{z}{r}|\psi|^2,d^3r$ & $\langle J_{p,z}\rangle$ & $\langle d_z\rangle$ \\\hline
$2$ & $3$ & $1,0,-1$ & $-\tfrac{1}{2},0,\tfrac{1}{2}$ &
$\hbar,0,-\hbar$ & $3d_B,0,-3d_B$\\
$3$ & $\tfrac{9}{2}$ & $2,1,0,-1,-2$ & $-\tfrac{2}{3},-\tfrac{1}{3},0,\tfrac{1}{3},\tfrac{2}{3}$ &
$2\hbar,\hbar,0,-\hbar,-2\hbar$ &
$\tfrac{9}{2}d_B,\tfrac{9}{4}d_B,0,-\tfrac{9}{4}d_B,-\tfrac{9}{2}d_B$
\label{TabI}
\end{tabular}
\end{center}
\end{table*}

\begin{figure}[t]
\includegraphics[width=1.0\columnwidth]{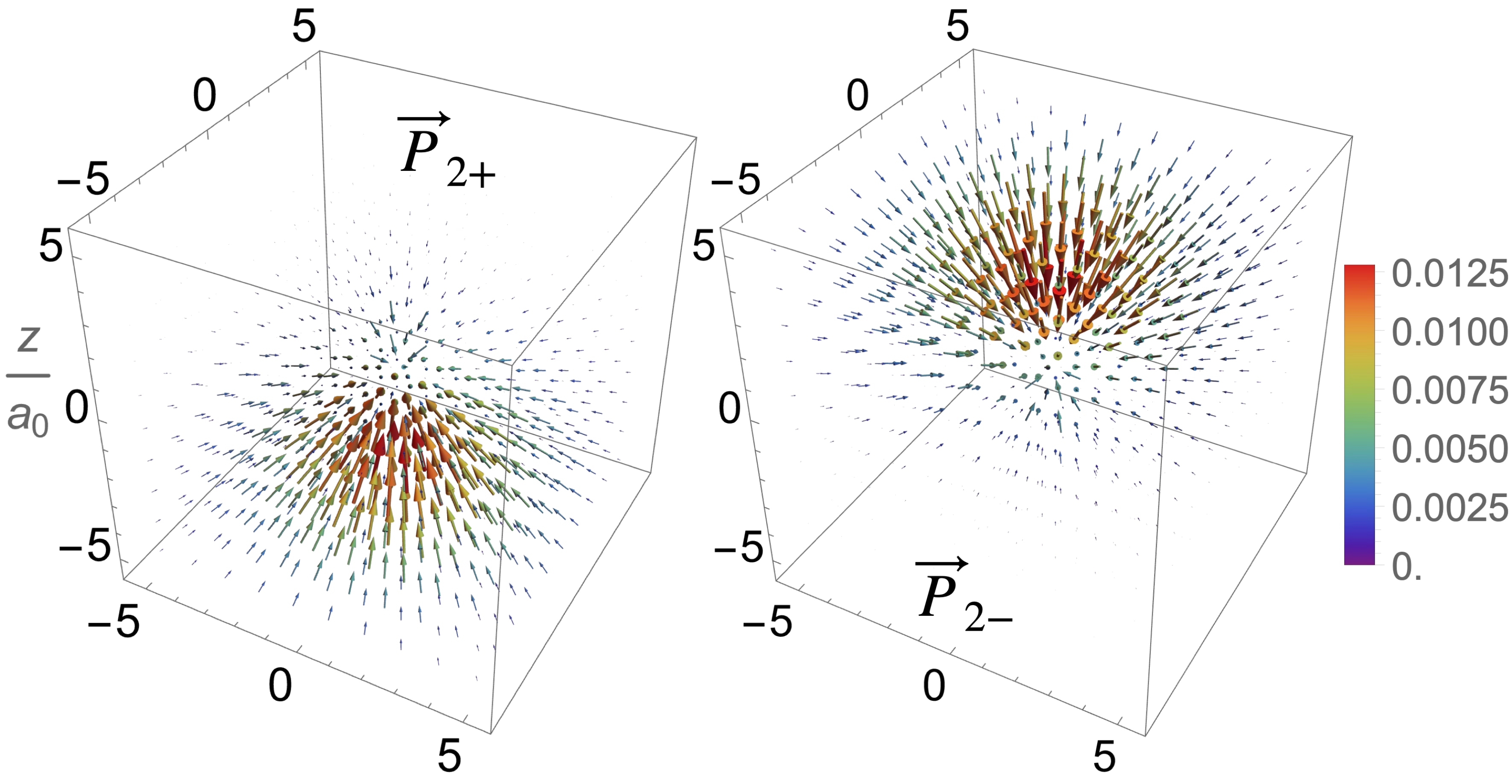}
\caption{Effective microscopic polarization vector-fields for $n=2$ Stark states.}
\label{Poln}
\end{figure}

Applying the dual Ohanian model introduced in Sec.~\ref{Oh}, we compute the \emph{internal} (relative-coordinate) electron charge density from the Stark wavefunctions. The hydrogen atom is a two-body system and separates into centre-of-mass and relative coordinates, $\Psi(\vec r_e,\vec r_p)=\Phi(\vec R)\psi(\vec r)$ with $\vec r=\vec r_e-\vec r_p$. The corresponding unconditional lab-frame one-particle density depends on the centre-of-mass state $\Phi(\vec R)$, where the centre-of-mass coordinate is $\vec R=(m_e\vec r_e+m_p\vec r_p)/(m_e+m_p)$. For a freely propagating atom with a delocalised centre-of-mass wavefunction, the unconditional lab-frame density becomes translationally invariant (uniform) and therefore does not encode the internal structure; it is not the appropriate quantity for visualising internal polarization fields. Throughout, we work in the internal (centre-of-mass/relative-coordinate) description, so that the associated microscopic ``effective'' fields are internal fields defined in the atomic rest frame. Considering this, we calculate the effective electron charge density directly from the internal wavefunction,
\begin{equation}
\begin{aligned}
\rho_{2\pm}(r,\theta) = -e\,|\psi_{2\pm}(r,\theta)|^2,
\end{aligned}
\label{eq:rho_internal}
\end{equation}
and thus the effective microscopic polarization from the charge density,
\begin{align}
&\vec{P}_{2\pm}(r,\theta)=r\,\rho_{2\pm}(r,\theta)\hat{r}= \\
&-\,\frac{e\,r}{8\pi}\Big[R_{20}^2 + 3 R_{21}^2\cos^2\theta \;\pm\; 2\sqrt{3}\,R_{20}R_{21}\cos\theta
\Big]\hat{r},
\label{eq:Pr_component}
\end{align}
which is plotted in Fig. \ref{Poln}.
\begin{figure}[t]
\includegraphics[width=1.0\columnwidth]{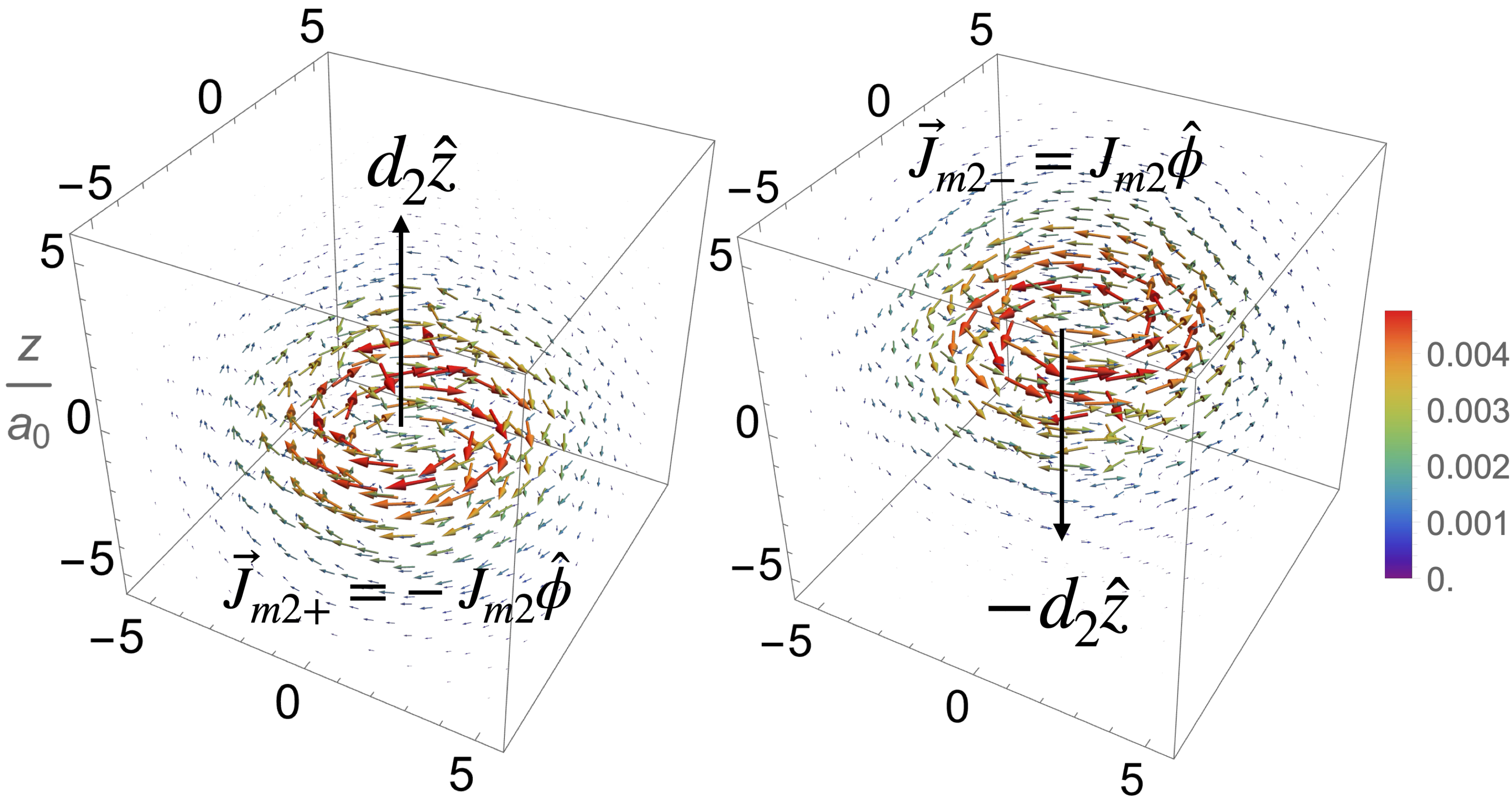}
\caption{Magnetic current vector-field ,$\vec{J}_{m2\pm}$ calculated from (\ref{Jm}), acting as a generator of the $\pm$ EDMs of magnitude $d_2$, via the left hand rule for the $n=2$ Stark states}
\label{MCurr}
\end{figure}
\begin{figure}[t]
\includegraphics[width=1.0\columnwidth]{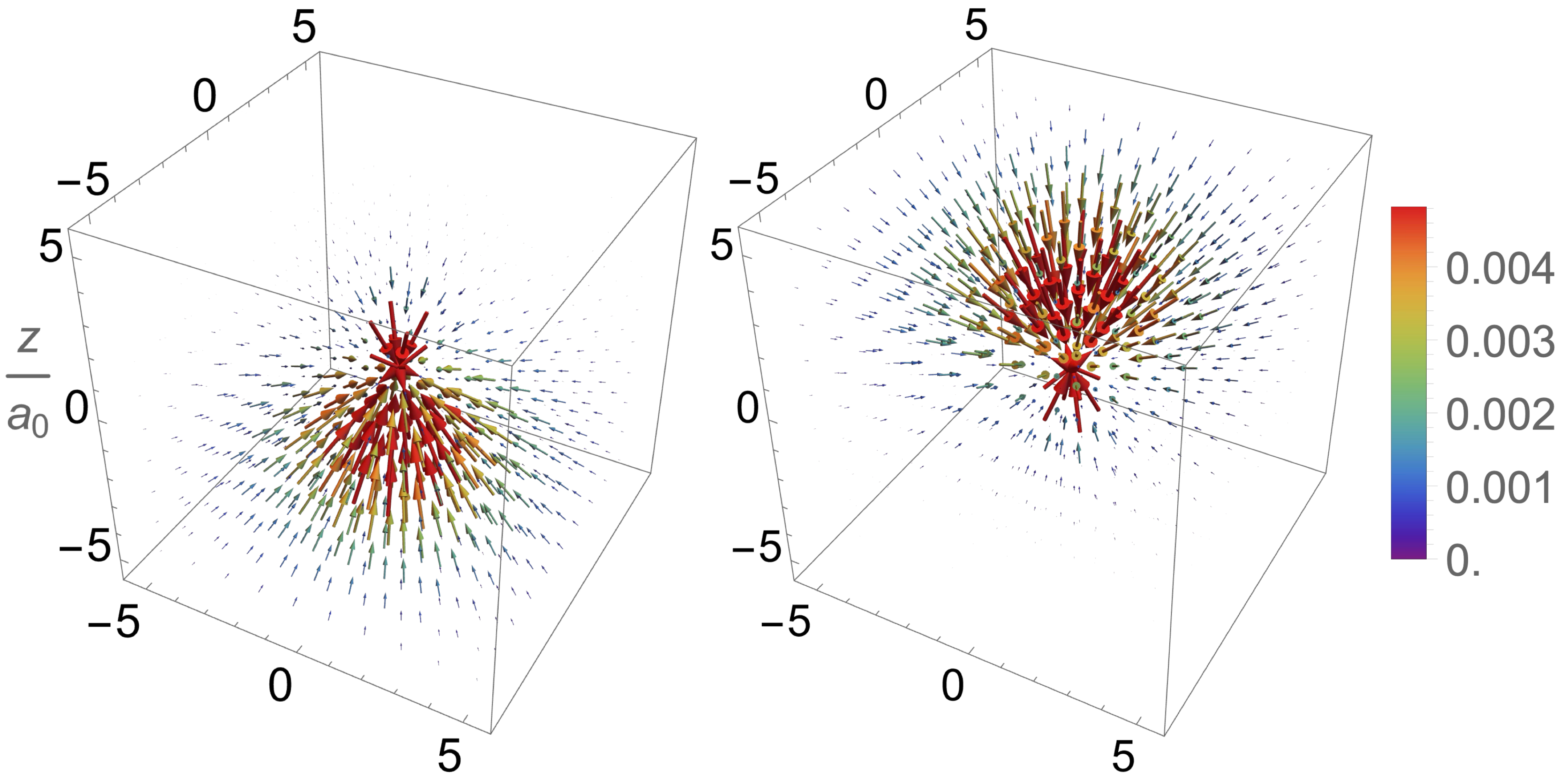}
\caption{The Runge-Lenz vector density, $\vec{\mathcal A}$, for the $n=2$ Stark states, given by (\ref{RLdensity}).}
\label{RLenz}
\end{figure}
\begin{figure}[t]
\includegraphics[width=1.0\columnwidth]{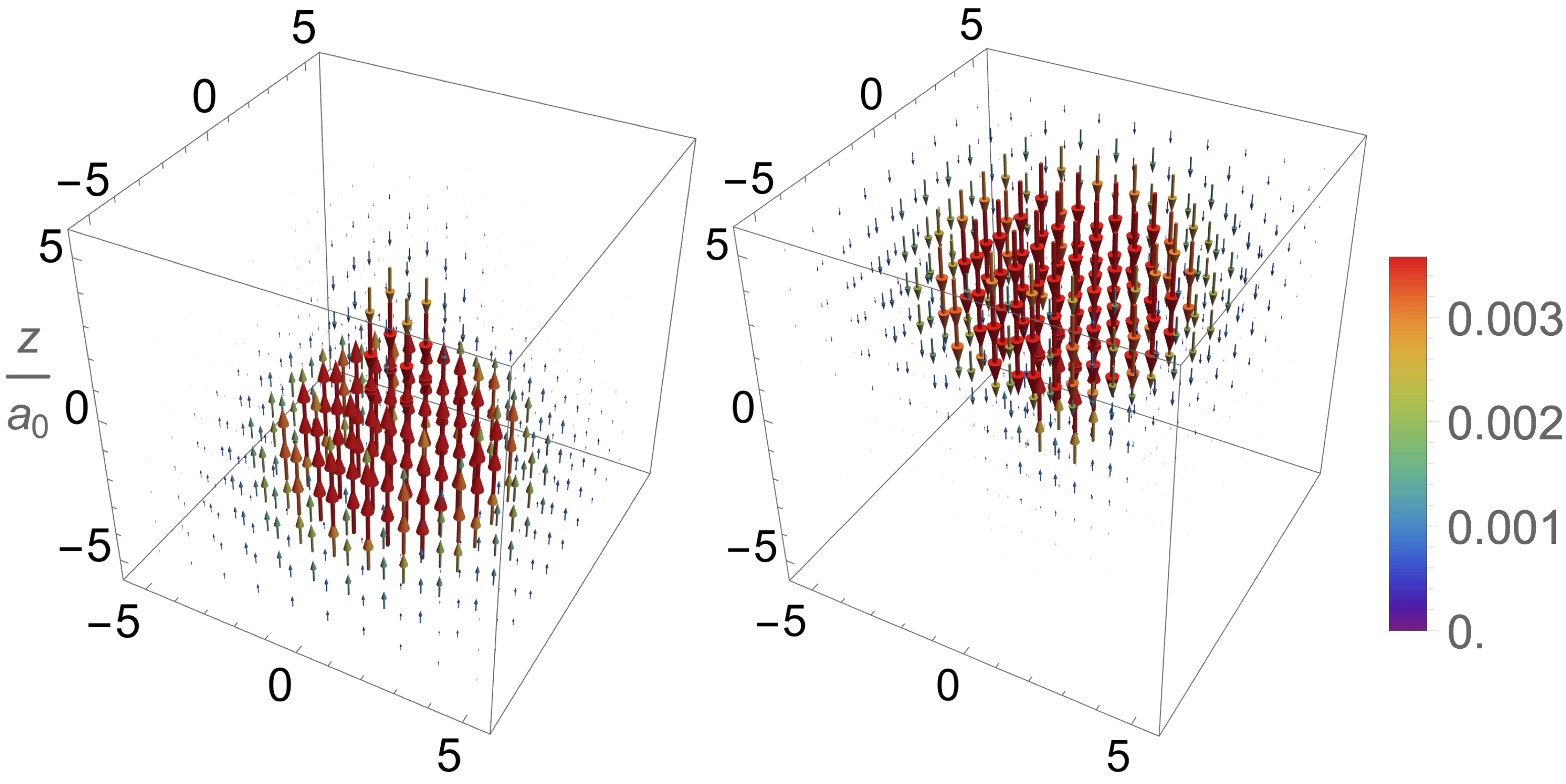}
\caption{The z-component of the Runge-Lenz vector density, $\vec{\mathcal A_z}$, for the $n=2$ Stark states, defined in (\ref{RLzdensity}).}
\label{RLz}
\end{figure}
Substituting (\ref{eq:Pr_component}) into (\ref{MCdens}), the effective magnetic current density may be calculated to be,
\begin{equation}
\begin{aligned}
&\vec{J}_{m2\pm}=-\frac{1}{\epsilon_0} \nabla \times \vec{P}_{2\pm}= \\
&-\frac{e \sin \theta R_{21}}{4 \pi \epsilon_0}\left[3 R_{21}(r) \cos \theta \pm \sqrt{3} R_{20}(r)\right] \hat{\phi}
\end{aligned}
\label{Jm}
\end{equation}
Note that $\vec{J}_{m2\pm}=\mp J_{m2}\,\hat{\phi}$ is purely azimuthal; it vanishes on the symmetry axis (where $\sin\theta=0$), and the $\pm$ sign reverses the direction between the two Stark states, as shown in Fig.~\ref{MCurr}. Furthermore, Eq.~(\ref{dPol2}) implies a left-hand rule relating the magnetic current to the dipole moment, $d_2$. Substituting Eq.~(\ref{Jm}) into Eq.~(\ref{dPol2}), one readily confirms $d_2=3d_B$.

\section{The Electric G-factor}

The hidden Runge-Lenz (Kepler) symmetry of the Coulomb problem (see Apendix \ref{RL}) explains why hydrogen shows a \emph{linear} Stark effect. To calculate expectation values, we can start from the Runge-Lenz density,
\begin{equation}
\begin{aligned}
\vec{\mathcal A}(\vec{r})
=&\frac{\hbar^2}{m}\,\mathrm{Im}\!\Big[(\vec{\nabla}\psi^{\ast})\times(\vec{r}\times\vec{\nabla}\psi)\Big]
-\kappa\,\frac{\vec{r}}{r}\,|\psi(\vec r)|^2, \\
&\int \vec{\mathcal A}\,d^3r=\langle \hat{\vec{A}}\rangle.
\label{RLdensity}
\end{aligned}
\end{equation}
 For the hydrogen Stark eigenstates we can choose $\psi$ real (no vector potential), so the first current-like term in (\ref{RLdensity}) vanishes and the  expectation value of the z-component of the Runge-Lenz vector becomes,
\begin{equation}
\langle A_z\rangle= \int \mathcal A_z\,d^3r
= -\,\kappa\int \frac{z}{r}\,|\psi(\vec{r})|^2\,d^3r,
\label{RLzdensity}
\end{equation}
and the scaled vector in units of angular momentum (see appnedix \ref{RL}), becomes
\begin{equation}
\langle A_{sc,z}\rangle= -\hbar n\int \frac{z}{r}\,|\psi(\vec{r})|^2\,d^3r=\langle\hat J_{p,z}\rangle,
\end{equation}
where we define $\hat J_{p,z}$ as the z-component pseudo angular momentum, $\hat J_{p}$, in parity space, which is essentially defined as the scaled Runge-Lenz vector. It follows that $\hat J_{p,z}$ is quantized in units of $\hbar$ so that,
\begin{equation}
 \langle J_{p,z}\rangle=\hbar\,k\ , k=n_1-n_2\in\{-n+1,\ldots,n-1\},
\end{equation}
so
\begin{equation}
 \int \frac{z}{r}\,|\psi(\vec r)|^2\,d^3r\ =\ -\,\frac{k}{n}.
 \label{DS}
\end{equation}
Using the analogous relation to the magnetic Landé factor defined by (\ref{ZM})-(\ref{mgf}),
\begin{equation}
\hat{\vec{d}} = \frac{g_Ed_B}{\hbar}\hat{\vec{J}}_p, 
\label{egf}
\end{equation}
where $g_E(n) = \frac{3}{2}n$ is the electric Landé factor and $d_B = e a_0$ from (\ref{BElec}) defines the Bohr dipole scale, the expectation value of the dipole moment along the field direction and the Stark energy level shift, $\Delta E_S $, becomes,
\begin{equation}
\begin{aligned}
&\langle d_z\rangle=\frac{g_E\,d_B}{\hbar}\,\langle J_{p,z}\rangle= g_E\,d_B\,k \\
&\Delta E_S = g_Ekd_BE_z
\end{aligned}
\end{equation}
These expressions show that the induced dipole moment and observed energy level shift are proportional to the projection quantum number $k$, scaled by the electric Landé factor and the fundamental dipole unit. Also, the density sum rule in (\ref{DS}) is a compact way to verify that the \emph{local} Runge-Lenz density reproduces the \emph{global} SU(2) quantization, giving a direct bridge to the electric Land\'e factor via (\ref{egf}). Just as magnetic $g$-factors differ between systems depending on their internal structure and dynamics, the electric $g$-factor encodes the efficiency with which the electron orbital in parity space is converted to a permanent polarization through the effective magnetic current loop.

\begin{table*}[t]
\caption{Duality between the magnetic Zeeman and electric Stark effects.}
\begin{center}
\begin{tabular}{lll}
\toprule
 & \textbf{Magnetic Sector (Zeeman)} & \textbf{Electric Sector (Linear Stark)} \\
\midrule
External field & $\vec{B}$ (pseudovector)& $\vec{E}$ (polar vector)\\
Coupled operator & Total angular momentum $\hat{\vec{J}}$ & Pseudo-angular momentum $\hat{\Vec{J}}_{p}$ \\
Conserved generators & z-component, $\hat{J}_z$ & z-component, Runge-Lenz component, $\hat{A}_{sc,z}$ \\
Bohr Model & $\mu _B=\frac{e\hbar}{2m_e}$ Bohr magneton & $d_B=\frac{2\mu_{\mathrm{B}}}{c\alpha}=e a_0$ Bohr EDM\\
Good quantum number & $m_J$ (magnetic) & $k$ (Stark or parabolic) \\
Dipole moment & $\hat{\vec{\mu}} = g_M\mu_B \hat{\vec{J}}/\hbar$ & $\hat{\vec{d}} = g_E d_B \hat{\vec{J}}_p/\hbar$ \\
Energy level Shift & $\Delta E_M = g_Mm_j\mu_BB_z$ & $\Delta E_S = g_Ekd_BE_z$ \\
Hamiltonian & $H_M = -\vec{\mu}\!\cdot\!\vec{B}$ & $H_S = -\vec{d}\!\cdot\!\vec{E}$ \\
Field coupling (total) & $\displaystyle H_M = -(\vec{\mu}_L + \vec{\mu}_S)\!\cdot\!\vec{B}$ & $\displaystyle H_S = -(\vec{d}_{orb} + \vec{d}_{e})\!\cdot\!\vec{E}$ \\
Linear response & Zeeman splitting & Stark splitting \\
Residual symmetry & $SO(4) \!\to\! SO(2)$ & $SO(4) \!\to\! SO(2)\!\times\! SO(2)$ \\
Parity Symmetry & Conserved (no mixing) & Broken, Mixed opposite parity states \\
Time Reversal Sym & Broken & Conserved \\
Basis states & $|\!\uparrow\rangle, |\!\downarrow\rangle$ & $|+\rangle, |-\rangle$ (even/odd parity) \\
Source & Circulating \emph{charge current} density & Circulating \emph{magnetic current} density \\
Intrinsic operator & $\displaystyle \hat{\vec S} = \frac{\hbar}{2}\vec{\sigma}$\ Spin operator  & $\displaystyle \hat{\vec S} = \frac{\hbar}{2}\vec{\sigma}$ Spin operator\\
Physical interpretation & Precession in $\vec{B}$; magnetic polarization & Polarization in $\vec{E}$; parity mixing \\
Geometric picture & Rotation in real space & Rotation in parity space \\
\bottomrule
\end{tabular}
\end{center}
\label{ZS}
\end{table*}

\subsection{Electric Spin}

To include an intrinsic dipole moment with the induced orbital effects one can combine (\ref{espin}) and (\ref{egf}) to obtain,
\begin{equation}
\hat{\vec d}_{tot}= d_B\big[g_E\hat{\vec J}_p\hbar + g_E^{e}\,\langle \hat{\vec S}\rangle/\hbar\big],~~g_E^{e}=\frac{2d_{\rm int}}{d_B}
\label{egftot}
\end{equation}
In this case the intrinsic EDM is alined with the particles spin, and is related to a probability magnetic current produced in position space by the spin operator. Spin is a pseudovector (axial vector), it behaves like an ordinary vector under rotations, but under spatial inversion (parity) it does not change sign, unlike a true polar vector. Under time reversal it does flip sign. Consequently, an intrinsic EDM is parity-odd and time-reversal-odd, so detecting a nonzero intrinsic EDM would signal CP violation.

In contrast, there exist many \emph{fermionic} quasiparticles whose two-component “spinor’’ resides in an \emph{internal} space and couples \emph{electrically}. Some canonical electric-sector realizations include graphene’s Dirac carriers, where the spinor acts on the A/B \emph{sublattice} degree of freedom and couples to inversion-breaking scalar potentials introduced by substrates or gating \cite{CastroNeto2009,Novoselov2005}; the closely related \emph{valley} pseudospin associated with the two inequivalent Brillouin–zone corners, which can be driven by electric fields through inversion asymmetry and Berry-curvature (valley Hall) effects \cite{Schaibley2016}; and bilayer graphene’s \emph{layer/orbital} pseudospin, controllable by a perpendicular displacement field (interlayer bias) that opens and tunes an energy gap \cite{McCann2006}. Cold-atom Fermi gases provide another electric route: two-level \emph{hyperfine} manifolds act as a tunable pseudospin-\(\tfrac{1}{2}\) coherently manipulated via optical (electric-dipole) Raman coupling and AC–Stark shifts \cite{Bloch2008}.

In this sense, the Stark/EDM pseudo-angular momentum $\hat{\vec{J}}_p$ is of the same kind: an \emph{orbital}, internal \(SU(2)\) (Runge-Lenz) that couples opposite-parity components within a fixed $n$ manifold, while the underlying particles remain bona fide fermions.

\section{Discussion and Implications}

The framework presented here provides a concrete route to visualizing electric dipole moments in close analogy to magnetic dipole moments (Tab.~\ref{ZS}). Introducing an effective magnetic current associated with permanent polarization yields a dual description of a permanent EDM and motivates an electric Bohr moment and electric $g$-factor. This electromagnetic-duality viewpoint highlights how internal circulations, whether represented as electric or magnetic currents, produce observable moments coupled to external fields, and how these couplings encode (and, in symmetry-violating cases, break) fundamental symmetries. The quantized boundary-current picture also connects naturally to modern polarization theory in materials, bridging Berry-phase descriptions and classical electromagnetic analogies.

A key point is that the pseudo-angular-momentum used here is general at the level of an internal two- (or few-) level subspace: whenever the dynamics is confined to an opposite-parity manifold coupled by an electric-dipole interaction, an $SU(2)$ pseudospin (Pauli-operator) algebra acting in that parity subspace naturally organizes Stark mixing and the associated induced dipoles. What is special to the hydrogenic Coulomb problem is the \emph{explicit realisation} of this $SU(2)$ generator in terms of continuous-space symmetry operators: within a fixed hydrogenic $n$-manifold the pseudo-angular momentum can be identified in closed form with a generator proportional to the scaled Runge-Lenz structure (via the hidden $SO(4)\simeq SU(2)\times SU(2)$ symmetry). Thus, the $SU(2)$ parity-subspace organisation is the general layer, while the Runge--Lenz proportionality provides the canonical hydrogenic realisation used here as an analytic benchmark.

This viewpoint mirrors standard classical impressed-source and field-equivalence formalisms \cite{RHbook2012,Balanis2012}: the driving EMF that separates charge may be represented as an impressed electric field, equivalently as a polarization field with nonzero curl. Because the EMF is localized, the boundary carries (i) dipolar charge that terminates the polarization and (ii) a fictitious magnetic surface current parallel to the polarization, reproducing external fields and terminal responses without modeling the internal workings of the source; this underpins our analyses of active classical dipoles using \emph{magnetic} equivalence currents \cite{TobarVP,Tobar2021}, and effective EMF generation in axion electrodynamics \cite{TobarModAx19,TOBAR2020}. The Stark-sector theory developed here is the quantum dual of this equivalence-current picture: starting from the wavefunction's microscopic polarization $\vec P(\vec r)$ we define an \emph{effective magnetic probability current} $\vec J_m=-\epsilon_0^{-1}\nabla\times\vec P$, providing a dual-current representation of the polarization texture.  Importantly, the induced EDM arises physically from charge displacement via parity (Stark) mixing; $\vec J_m$ is not an additional causal mechanism, but a convenient field-equivalent description of the same internal structure. In this sense, the dual-Ohanian formulation makes the classical--quantum correspondence explicit: an impressed electric excitation is represented by a magnetic equivalence current in the classical dipole \cite{TobarVP,Tobar2021}, while Stark mixing is represented by the magnetic probability-current \emph{representation} of the quantum dipole.

\section*{Appendices}
\appendix

\section{Runge-Lenz Operator}
\label{RL}

The unperturbed hydrogen atom is described by the Coulomb Hamiltonian
\begin{align}
  \hat H_0 = \frac{\hat{\vec p}^2}{2\mu} - \frac{\kappa}{\hat r},
  \qquad
  \mu = \frac{m_e m_p}{m_e + m_p}\ \ [\mathrm{kg}],
\end{align}
where $\mu$ is the reduced mass of the electron-proton system. The Coulomb coupling constant is given by
\begin{align}
  \kappa = \frac{e^2}{4\pi\epsilon_0}\ \ [\mathrm{J\,m}],
  \qquad
  e\ \ [\mathrm{C}],\quad
  \epsilon_0\ \ [\mathrm{F\,m^{-1}}],
\end{align}
and the characteristic Bohr radius follows as
\begin{align}
  a_0 = \frac{4\pi\epsilon_0\hbar^2}{\mu e^2}\ \ [\mathrm{m}],
  \qquad
  \hbar\ \ [\mathrm{J\,s}].
\end{align}

The canonical position and momentum operators are defined as
\begin{align}
  \hat{\vec r} = (\hat x, \hat y, \hat z), \quad
  \hat r = |\hat{\vec r}|, \quad
  \widehat{\vec r} = \frac{\hat{\vec r}}{\hat r},
\end{align}
\begin{align}
  \hat{\vec p} = -\,i\hbar\,\vec{\nabla}\ \ [\mathrm{kg\,m\,s^{-1}}],
\end{align}
and the orbital angular momentum operator takes the standard form
\begin{align}
  \hat{\vec L} = \hat{\vec r}\times\hat{\vec p}\ \ [\mathrm{J\,s}].
\end{align}

A less familiar but equally fundamental conserved quantity of the Coulomb problem is the Runge-Lenz vector \cite{Hughes1967,PhysRev.133.B1352,atoms11070105,sym12101719}, defined (in symmetrized operator form) as
\begin{align}
  \hat{\vec A}
  = \frac{1}{2\mu}\big(\hat{\vec p}\times\hat{\vec L} - \hat{\vec L}\times\hat{\vec p}\big)
  - \kappa\,\frac{\widehat{\vec r}}{r}\ \ [\mathrm{J\,m}].
\end{align}
This operator generates a hidden dynamical symmetry of the hydrogen atom, corresponding to the conserved direction of the elliptical orbit in the classical limit. Its magnitude and commutation relations with $\hat{\vec L}$ lead to the well-known $SO(4)\!\simeq\!SU(2)\!\times\!SU(2)$ symmetry of the bound-state spectrum \cite{Hughes1967}.

To obtain an angular-momentum-dimension, we define a scaled Runge-Lenz operator,
\begin{align}
  \hat{\vec A}_{\mathrm{sc}} \equiv \frac{\sqrt{\mu}\hat{\vec A}}{\sqrt{-2\hat H_0}}\ \ [\mathrm{J\,s}],
  \qquad
  \hat{\vec J}_p \equiv \hat{\vec A}_{\mathrm{sc}},
  \label{eq:AA_sc}
\end{align}
which introduces a pseudo-angular momentum $\hat{\vec J}_p$ with the same algebraic structure as $\hat{\vec L}$.

On a hydrogenic energy eigenspace $\hat H_0|_{E_n} = E_n|_{E_n}$ with
\begin{align}
  E_n = -\,\frac{\mu \kappa^2}{2\hbar^2 n^2},
\end{align}
the scaling factor becomes a number:
\begin{align}
  \sqrt{-2E_n/\mu} = \frac{\kappa}{\hbar n} 
  \Rightarrow 
  \hat{\vec A}_{\mathrm{sc}}
  = \frac{\hbar n}{\kappa}\,\hat{\vec A}
  =\frac{n}{c \alpha}\,\hat{\vec A}
\end{align}
 on the $n$ manifold. Thus, within each principal quantum manifold $n$, the scaled Runge-Lenz vector plays the role of a conserved pseudo-angular momentum, forming with $\hat{\vec L}$ the generators of the $SU(2)\times SU(2)$ symmetry that underlies the degeneracy of the hydrogen spectrum.

\section{Hydrogen quantum numbers (spherical vs. parabolic)}

In hydrogen, the electron’s orbital quantum numbers are the principal quantum number $n=1,2,3,\ldots$, the orbital angular momentum $\ell=0,1,\ldots,n-1$ (with spectroscopic labels $s,p,d,f,\ldots$ for $\ell=0,1,2,3,\ldots$), and the magnetic quantum number $m=-\ell,-\ell+1,\ldots,\ell$, which set $L^2=\ell(\ell+1)\hbar^2$ and $L_z=m\hbar$. In an external electric field it is convenient to use the \emph{parabolic} set $(n_1,n_2,m)$, obtained by separating the Schr\"odinger equation in parabolic coordinates $(\xi=r+z,\ \eta=r-z,\ \phi)$. Here $n_1,n_2=0,1,2,\ldots$ are \emph{not} principal quantum numbers: they count radial-like nodes along the two parabolic directions and, together with $m$, obey the constraint $n_1+n_2+|m|+1=n$ (where $n$ remains the single principal quantum number). The combination $k\equiv n_1-n_2\in\{-(n-1),\ldots,n-1\}$ labels the linear Stark splitting and controls the expectation values $\langle z\rangle=\tfrac{3}{2}\,n\,a_0\,k$ and $\langle A_{\mathrm{sc},z}\rangle=\hbar k$. In zero field, the parabolic basis is just another unitary choice within the $n$ manifold, equivalent to the spherical $\lvert n\ell m\rangle$ basis.

\section{Hydrogenic $SO(4)$ structure and the pseudo-angular momentum}

For a hydrogenic atom/ion, the following familiar Coulomb-problem commutation relations encode the hidden $SO(4)$ symmetry of the bound-state ($H<0$) spectrum.
\begin{align}
[L_i,L_j] &= i\hbar\,\epsilon_{ijk}L_k, \label{eq:LL_comm}\\
[L_i,A_j] &= i\hbar\,\epsilon_{ijk}A_k, \label{eq:LA_comm}\\
[A_i,A_j] &= -2\mu H\, i\hbar\,\epsilon_{ijk}L_k, 
\label{eq:AA_comm}
\end{align}
For bound states it is convenient to use the scaled Runge--Lenz operator \eqref{eq:AA_sc} , so \eqref{eq:AA_comm} implies
\begin{equation}
[A_{{\rm sc},i},A_{{\rm sc},j}] \;=\; i\hbar\,\epsilon_{ijk}L_k,
\label{eq:AscAsc_comm}
\end{equation}
and $\vec A_{\rm sc}$ has the dimensions of angular momentum (like $\vec L$). One may then form the standard
$SO(4)\simeq SU(2)\times SU(2)$ generators
\begin{equation}
\vec J_{\pm} \;\equiv\; \frac{1}{2}\left(\vec L \pm \vec A_{\rm sc}\right),
\label{eq:Jpm_def}
\end{equation}
which satisfy the $SU(2)$ algebras
\begin{align}
[J_{\pm,i},J_{\pm,j}] &= i\hbar\,\epsilon_{ijk}J_{\pm,k}, \label{eq:Jpm_SU2}\\
[J_{+,i},J_{-,j}] &= 0. \label{eq:JpJm_commute}
\end{align}
Within a fixed $n$-manifold (degenerate subspace), these relations provide the rigorous $SU(2)$ structure
underpinning the Land\'e-factor analogy used below.

In the Stark/parity (two- or few-level) subspace relevant for the orbital EDM, we identify the pseudo-angular momentum $\vec J_p$ with the appropriate projected $SU(2)$ generator (equivalently, the corresponding component of $\vec J_{\pm}$ along the field axis) that labels the parabolic/Stark structure. Consequently, $\vec J_p$ obeys the standard $SU(2)$ commutation relations,
\begin{equation}
[J_{p,i},J_{p,j}] \;=\; i\hbar\,\epsilon_{ijk}J_{p,k},
\label{eq:Jp_SU2}
\end{equation}
justifying the Land\'e-factor form
$\hat{\vec d}_{\rm orb}=g_E d_B\,\hat{\vec J}_p/\hbar$ used in this work.

\begin{acknowledgments}
This work was funded by the ARC Centre of Excellence for Dark Matter Particle Physics, CE200100008.
\end{acknowledgments}


\begin{thebibliography}{10}

\bibitem{Stark1914}
J.~Stark.
\newblock Observations of the effect of the electric field on spectral lines i.
  transverse effect.
\newblock {\em Annalen der Physik}, 43:965--983, 1914.

\bibitem{Feynman57}
Richard~P. Feynman, Frank~L. Vernon, and Robert~W. Hellwarth.
\newblock Geometrical representation of the schr{\"o}dinger equation for
  solving maser problems.
\newblock {\em Journal of Applied Physics}, 28(1):49--52, 1957.

\bibitem{EDMBook}
Iosif~B. Khriplovich and Steve~K. Lamoreaux.
\newblock {\em CP Violation Without Strangeness, Electric Dipole Moments of
  Particles, Atoms, and Molecules}.
\newblock Texts and Monographs in Physics. Springer-Verlag Berlin Heidelberg,
  1997.

\bibitem{Sandars1965}
P.G.H. Sandars.
\newblock The electric dipole moment of an atom.
\newblock {\em Physics Letters}, 14(3):194--196, 1965.

\bibitem{Engel2013}
Jonathan Engel, Michael~J. Ramsey-Musolf, and U.~{van Kolck}.
\newblock Electric dipole moments of nucleons, nuclei, and atoms: The standard
  model and beyond.
\newblock {\em Progress in Particle and Nuclear Physics}, 71:21--74, 2013.
\newblock Fundamental Symmetries in the Era of the LHC.

\bibitem{DeMille2017}
David DeMille, John~M. Doyle, and Alexander~O. Sushkov.
\newblock Probing the frontiers of particle physics with tabletop-scale
  experiments.
\newblock {\em Science}, 357(6355):990--994, 2017.

\bibitem{Chupp2019}
T.~E. Chupp, P.~Fierlinger, M.~J. Ramsey-Musolf, and J.~T. Singh.
\newblock Electric dipole moments of atoms, molecules, nuclei, and particles.
\newblock {\em Rev. Mod. Phys.}, 91:015001, Jan 2019.

\bibitem{Regan2002}
B.~C. Regan, Eugene~D. Commins, Christian~J. Schmidt, and David DeMille.
\newblock New limit on the electron electric dipole moment.
\newblock {\em Phys. Rev. Lett.}, 88:071805, Feb 2002.

\bibitem{Resta1994}
Raffaele Resta.
\newblock Macroscopic polarization in crystalline dielectrics: the geometric
  phase approach.
\newblock {\em Rev. Mod. Phys.}, 66:899--915, Jul 1994.

\bibitem{Vanderbilt2018}
David Vanderbilt.
\newblock {\em Berry Phase in Electronic Structure Theory, Electric
  Polarization, Orbital Magnetization and Topological Insulators}.
\newblock Cambridge University Press, 2018.

\bibitem{Song2021}
Xue-Yang Song, Yin-Chen He, Ashvin Vishwanath, and Chong Wang.
\newblock Electric polarization as a nonquantized topological response and
  boundary luttinger theorem.
\newblock {\em Phys. Rev. Research}, 3:023011, Apr 2021.

\bibitem{TobarVP}
Michael~E. Tobar, Raymond~Y. Chiao, and Maxim Goryachev.
\newblock Active electric dipole energy sources: Transduction via electric
  scalar and vector potentials.
\newblock {\em Sensors}, 22(18), 2022.

\bibitem{Tobar2021}
Michael~E. Tobar, Ben~T. McAllister, and Maxim Goryachev.
\newblock Electrodynamics of free- and bound-charge electricity generators
  using impressed sources.
\newblock {\em Phys. Rev. Applied}, 15:014007, Jan 2021.

\bibitem{TobarModAx19}
Michael~E. Tobar, Ben~T. McAllister, and Maxim Goryachev.
\newblock Modified axion electrodynamics as impressed electromagnetic sources
  through oscillating background polarization and magnetization.
\newblock {\em Physics of the Dark Universe}, 26:100339, 2019.

\bibitem{Hwang2012}
H.~Y. Hwang, Y.~Iwasa, M.~Kawasaki, B.~Keimer, N.~Nagaosa, and Y.~Tokura.
\newblock Emergent phenomena at oxide interfaces.
\newblock {\em Nature Materials}, 11(2):103--113, 2012.

\bibitem{Liu2020}
X.Z. Liu, Z.X. Tang, Q.H. Li, Q.H. Zhang, X.Q. Yu, and L.~Gu.
\newblock Symmetry-induced emergent electrochemical properties for rechargeable
  batteries.
\newblock {\em Cell Reports Physical Science}, 1(6):100066, 2020.

\bibitem{Ohanian86}
Hans~C. Ohanian.
\newblock What is spin?
\newblock {\em American Journal of Physics}, 54(6):500--505, 06 1986.

\bibitem{Roussy23}
Tanya~S. Roussy, Luke Caldwell, Trevor Wright, William~B. Cairncross, Yuval
  Shagam, Kia~Boon Ng, Noah Schlossberger, Sun~Yool Park, Anzhou Wang, Jun Ye,
  and Eric~A. Cornell.
\newblock An improved bound on the electron's electric dipole moment.
\newblock {\em Science}, 381(6653):46--50, 2023.

\bibitem{Abel2020}
C.~Abel, S.~Afach, N.~J. Ayres, C.~A. Baker, G.~Ban, G.~Bison, K.~Bodek,
  V.~Bondar, M.~Burghoff, E.~Chanel, Z.~Chowdhuri, P.-J. Chiu, B.~Clement,
  C.~B. Crawford, M.~Daum, S.~Emmenegger, L.~Ferraris-Bouchez, M.~Fertl,
  P.~Flaux, B.~Franke, A.~Fratangelo, P.~Geltenbort, K.~Green, W.~C. Griffith,
  M.~van~der Grinten, Z.~D. Gruji\ifmmode~\acute{c}\else \'{c}\fi{}, P.~G.
  Harris, L.~Hayen, W.~Heil, R.~Henneck, V.~H\'elaine, N.~Hild, Z.~Hodge,
  M.~Horras, P.~Iaydjiev, S.~N. Ivanov, M.~Kasprzak, Y.~Kermaidic, K.~Kirch,
  A.~Knecht, P.~Knowles, H.-C. Koch, P.~A. Koss, S.~Komposch, A.~Kozela,
  A.~Kraft, J.~Krempel, M.~Ku\ifmmode~\acute{z}\else \'{z}\fi{}niak, B.~Lauss,
  T.~Lefort, Y.~Lemi\`ere, A.~Leredde, P.~Mohanmurthy, A.~Mtchedlishvili,
  M.~Musgrave, O.~Naviliat-Cuncic, D.~Pais, F.~M. Piegsa, E.~Pierre, G.~Pignol,
  C.~Plonka-Spehr, P.~N. Prashanth, G.~Qu\'em\'ener, M.~Rawlik, D.~Rebreyend,
  I.~Rien\"acker, D.~Ries, S.~Roccia, G.~Rogel, D.~Rozpedzik, A.~Schnabel,
  P.~Schmidt-Wellenburg, N.~Severijns, D.~Shiers, R.~Tavakoli~Dinani, J.~A.
  Thorne, R.~Virot, J.~Voigt, A.~Weis, E.~Wursten, G.~Wyszynski, J.~Zejma,
  J.~Zenner, and G.~Zsigmond.
\newblock Measurement of the permanent electric dipole moment of the neutron.
\newblock {\em Phys. Rev. Lett.}, 124:081803, Feb 2020.

\bibitem{CastroNeto2009}
A.~H. Castro~Neto, F.~Guinea, N.~M.~R. Peres, K.~S. Novoselov, and A.~K. Geim.
\newblock The electronic properties of graphene.
\newblock {\em Reviews of Modern Physics}, 81(1):109--162, 2009.

\bibitem{Novoselov2005}
K.~S. Novoselov, A.~K. Geim, S.~V. Morozov, D.~Jiang, M.~I. Katsnelson, I.~V.
  Grigorieva, S.~V. Dubonos, and A.~A. Firsov.
\newblock Two-dimensional gas of massless {Dirac} fermions in graphene.
\newblock {\em Nature}, 438:197--200, 2005.

\bibitem{Schaibley2016}
John~R. Schaibley, Hongyi Yu, Genevieve Clark, Pasqual Rivera, Jason~S. Ross,
  Kyle~L. Seyler, Wang Yao, and Xiaodong Xu.
\newblock Valleytronics in 2d materials.
\newblock {\em Nature Reviews Materials}, 1:16055, 2016.

\bibitem{McCann2006}
Edward McCann and Vladimir~I. Fal'ko.
\newblock Landau-level degeneracy and quantum hall effect in a graphite
  bilayer.
\newblock {\em Physical Review Letters}, 96:086805, 2006.

\bibitem{Bloch2008}
Immanuel Bloch, Jean Dalibard, and Wilhelm Zwerger.
\newblock Many-body physics with ultracold gases.
\newblock {\em Reviews of Modern Physics}, 80(3):885--964, 2008.

\bibitem{RHbook2012}
Roger~E. Harrington.
\newblock {\em Introduction to Electromagnetic Engineering}.
\newblock Dover Publications, Inc., 31 East 2nd Street, Mineola, NY 11501, 2nd
  edition, 2012.

\bibitem{Balanis2012}
Constantine~A Balanis.
\newblock {\em Advanced Engineering Electromagnetics}.
\newblock John Wiley,, 2012.

\bibitem{TOBAR2020}
Michael~E. Tobar, Ben~T. McAllister, and Maxim Goryachev.
\newblock Broadband electrical action sensing techniques with conducting wires
  for low-mass dark matter axion detection.
\newblock {\em Physics of the Dark Universe}, 30:100624, 2020.

\bibitem{Hughes1967}
J~W~B Hughes.
\newblock Stark states and {O}(4) symmetry of hydrogenic atoms.
\newblock {\em Proceedings of the Physical Society}, 91(4):810, 1967.

\bibitem{PhysRev.133.B1352}
Peter~J. Redmond.
\newblock Generalization of the runge-lenz vector in the presence of an
  electric field.
\newblock {\em Phys. Rev.}, 133:B1352--B1353, Mar 1964.

\bibitem{atoms11070105}
Jean-Christophe Pain.
\newblock On invariant vectors in the presence of electric and magnetic fields.
\newblock {\em Atoms}, 11(7), 2023.

\bibitem{sym12101719}
Marisol Berm{\'u}dez-Monta{\~n}a, Marisol Rodr{\'\i}guez-Arcos, Renato Lemus,
  Jos{\'e}~M. Arias, Joaqu{\'\i}n G{\'o}mez-Camacho, and Emilio Orgaz.
\newblock Algebraic dvr approaches applied to describe the stark effect.
\newblock {\em Symmetry}, 12(10), 2020.

\end{thebibliography}
\end{document}